# Advancing Specific Local Wine Identification through Integration of ICP-MS Mineral Wine Profiles and Machine Learning: A Case Study on Discrimination Among the Ten Beaujolais Crus

Yohann Clément[1*], Pierre Lantéri[1], Agnès Hagège[1], Bertrand Chatelet[5]*, Théodore Tillement[4], Coraline Duroux[4], Leticia Sarlo[4], François Lux[2,3], and Olivier Tillement[2]

[1]Université Claude Bernard Lyon 1, CNRS, ISA (Institut des Sciences Analytiques), UMR 5280, Villeurbanne, France

[2]Institut Lumière-Matière, UMR 5306, Université Lyon1-CNRS, Université de Lyon, Villeurbanne Cedex, France

[3]Institut Universitaire de France (IUF), 75231 Paris, France.

[4]M&Wine, Villeurbanne, France

[5]SICAREX Beaujolais, Villeranche sur Saône, France

*Corresponding author: yohann.clement@univ-lyon1.fr

## Abstract

The Beaujolais vineyard, located in France's Auvergne–Rhône–Alpes region, is internationally recognized for its geological diversity and the strong identity of its ten crus. Guaranteeing authenticity of these appellations is essential for producers and regulators, as mislabeling and fraud can undermine both economic value and consumer trust.

In this study, we introduce the Mineral Wine Profile (MWP), a high-throughput ICP-MS protocol enabling the semi-quantitative analysis of more than 40 mineral elements. A total of 1 288 Gamay wines from the 2022 and 2023 vintages, collected across the ten Beaujolais crus, were analyzed. Multivariate statistical methods and machine learning were applied to assess their discriminatory power. Principal Component Analysis clearly separated vintages, while Random Forest models achieved 98% accuracy for vintage classification. Mixture Discriminant Analysis further differentiated the ten crus with an average accuracy of 92%. Misclassifications mostly occurred between geographically neighboring crus, reflecting overlapping soil structures and terroir continuity. Element–soil correlations (e.g., Fe enriched in clay-rich soils, Mn in sandy soils, Cs in silty soils) provided mechanistic insights into terroir differentiation. These results confirm that wine mineral fingerprints can act as reliable chemical memories of origin.

By combining elemental profiling with Machine Learning, this work demonstrates a robust, reproducible, and scalable approach for intra-regional wine authentication. Beyond its scientific contribution, the MWP approach offers practical applications for the valorization of terroirs, protection of Appellations d'Origine Contrôlée (AOC), and fraud prevention in the wine industry.

## Introduction

The Beaujolais vineyard, located in the Auvergne-Rhône-Alpes region of France, is a viticultural area of international reputation producing primarily red wines from the Gamay grape variety. While Beaujolais Nouveau has achieved worldwide recognition, the ten Beaujolais crus (Saint-Amour, Juliénas, Chénas, Moulin-à-Vent, Fleurie, Chiroubles, Morgon, Régnié, Côte de Brouilly, and Brouilly) represent the highest added value of the region. These crus are associated with greater ageing potential, stronger typicity, and market prices up to several times higher than those of generic Beaujolais wines. This economic importance makes authenticity verification of Beaujolais crus a key issue for both producers and consumers, particularly in international trade where mislabeling and fraud remain risks.

From a geological perspective, Beaujolais is characterized by remarkable diversity, with formations including granite, schist, clay-limestone, sandy, and silty soils[1], [2], [3], [4], [5], [6], [7], [8]. These soils vary in mineralogical composition, water retention capacity, and fertility, directly influencing vine nutrition and grape composition. They can be broadly categorized as follows:

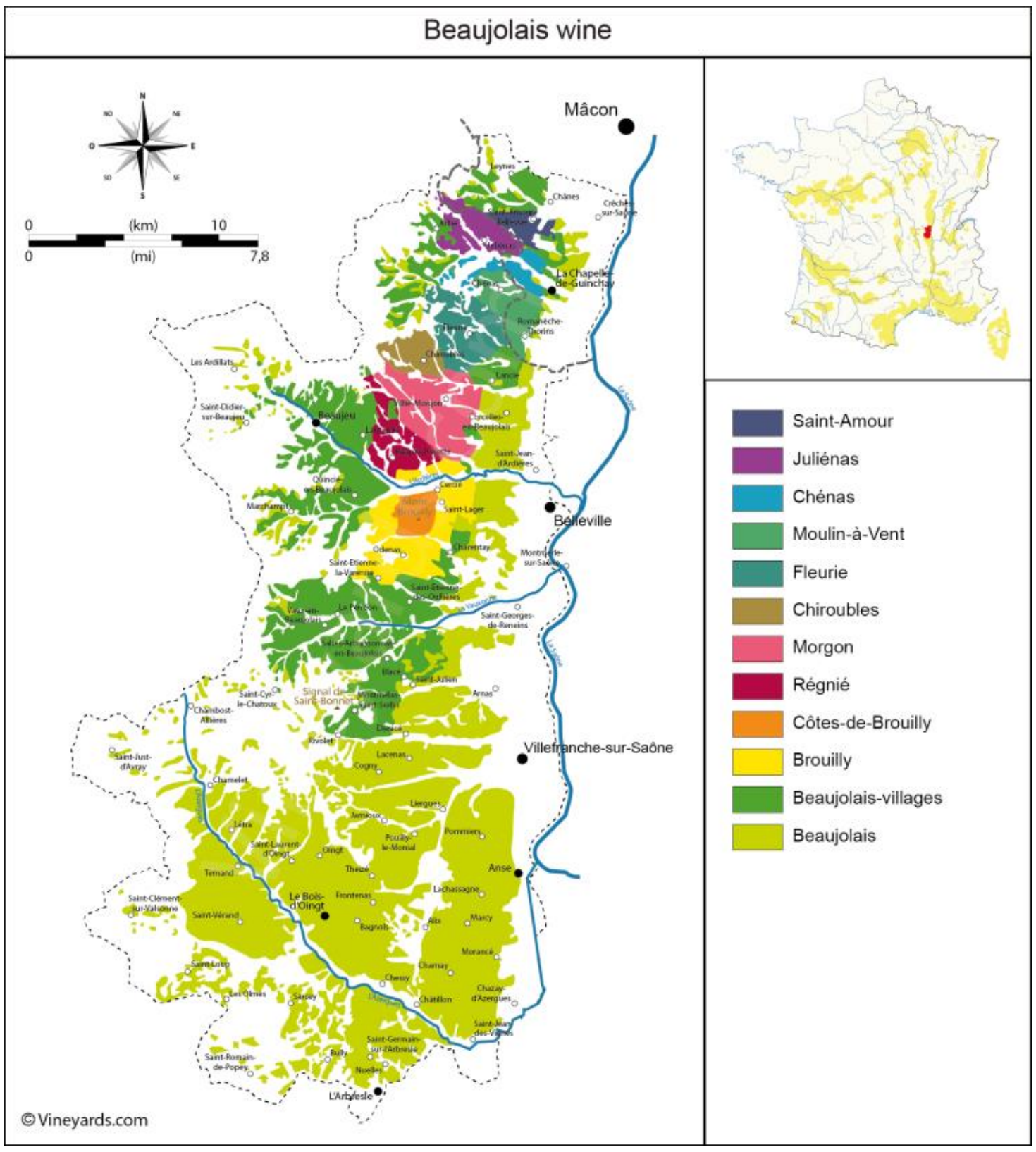

**Fig. 1: Maps form Beaujolais by Vineyards.com; Crus of Beaujolais AOP: Saint-Amour, Juliénas, Chénas, Moulin à Vent, Fleurie, Chiroubles, Morgon, Régnié, Côtes de Brouilly and Brouilly**

- ***Clay:*** Composed of clay and clay-limestone mixtures, these soils are rich in minerals and cations ($Ca^{2+}$, $Mg^{2+}$, $K^{+}$). They retain water efficiently, ensuring regular vine supply even during dry periods, while also conferring high buffering capacity. Wines from clay-rich soils are typically structured, with firm tannins, more concentrated color, and greater ageing potential.
- ***Sandy:*** Characterized by high permeability and low water retention, sandy soils provide rapid drainage but limited nutrient availability. These conditions result in lighter vine vigor, earlier grape ripening, and wines that are generally fresher, more aromatic, and fruit-driven, though with less structural depth. Sandy terroirs are often associated with styles suitable for early consumption, including Beaujolais primeurs.
- ***Silt:*** Composed of fine particles with high water retention capacity, silty soils act as important water reservoirs during summer droughts. However, they can lead to excessive vigor if not managed carefully, requiring specific viticultural practices (e.g., canopy control, yield regulation). Wines from silty soils are frequently balanced, with a round mouthfeel and supple tannins, reflecting the buffering effect of this soil type.

This heterogeneity of soil types contributes to the diversity of wine styles observed among the ten crus, despite the use of a single grape variety. It provides the scientific basis for investigating the relationship between mineral composition and terroir identity.

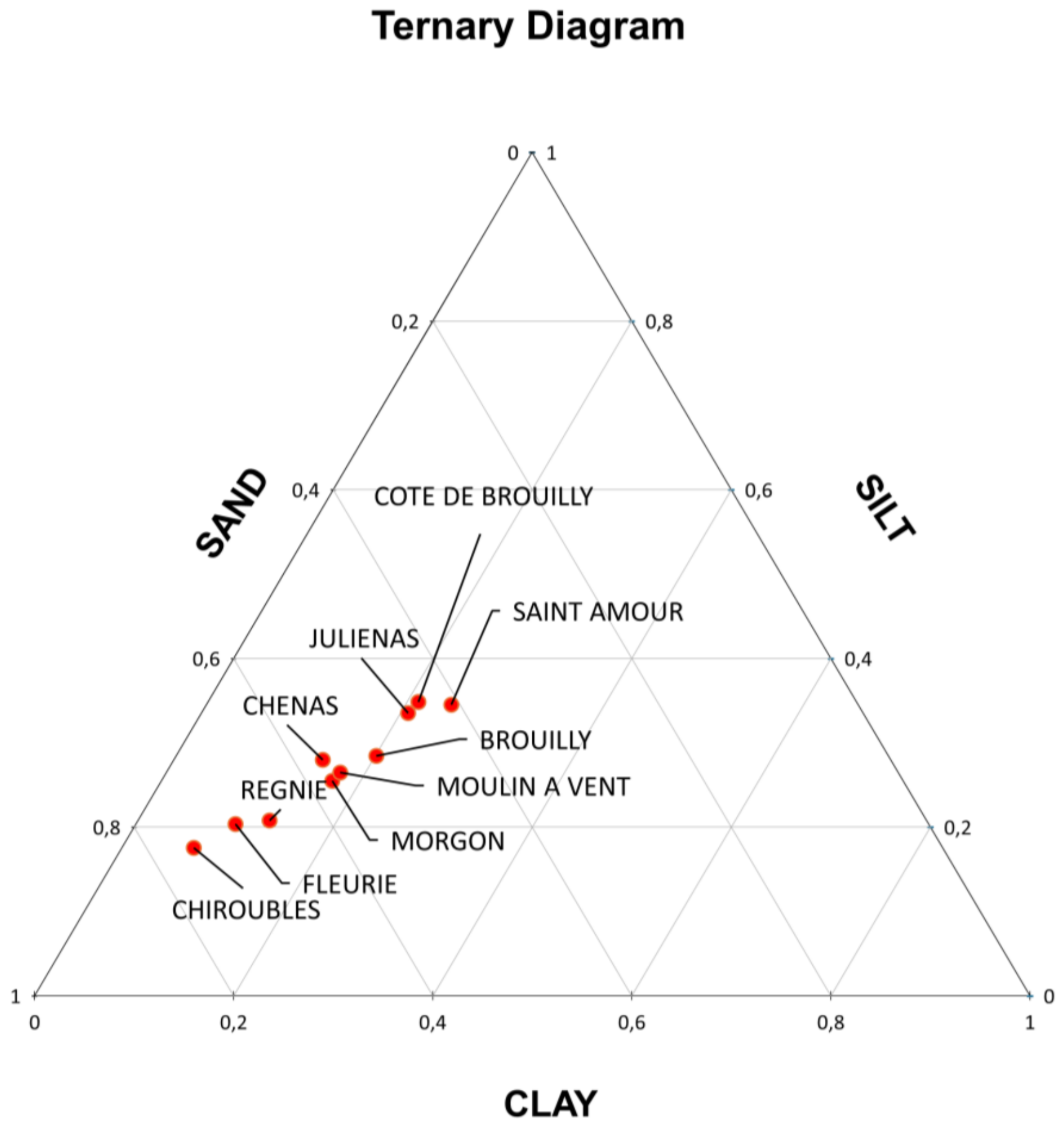

**Fig. 2 : Ternary diagram illustrating soil composition analysis of the 10 Beaujolais crus : Saint-Amour, Juliénas, Chénas, Moulin à Vent, Fleurie, Chiroubles, Morgon, Régnié, Côtes de Brouilly, and Brouilly**

The mineral composition of wines reflects vineyard soils, environmental conditions, vineyard management, and vinification processes. It can therefore be considered a chemical memory of origin[9], [10], [11], [12], [13], [14]. Over the past two decades, several studies have applied ICP-MS or related elemental techniques to differentiate wines by origin [8], [11], [13], [15], [16], [17], [18], [19]. These studies demonstrated that multielemental fingerprints, combined with chemometrics, are effective for distinguishing wines at regional or national levels. However, most existing works have two main limitations:

- They focus on inter-regional classification (e.g., Bordeaux vs Burgundy) rather than intra-regional scales such as Beaujolais crus.
- They rely on relatively small sample sizes, limiting the generalizability of their models.

To address these limitations, we introduce the Mineral Wine Profile (MWP), a semi-quantitative ICP-MS protocol enabling high-throughput analysis of more than 41 mineral elements (>200 samples/day). The MWP facilitates the construction of robust databases, which can then be exploited with Machine Learning (ML) and machine learning techniques.

Previous studies have successfully applied ICP-MS and chemometric tools for inter-regional classification of wines [11], [18], [20], [21]. However, these works were mostly limited in sample size (<300) and addressed large-scale geographical contrasts (e.g., Bordeaux vs Burgundy). To our knowledge, no study has yet demonstrated stable performance across resampling discrimination among the ten Beaujolais crus using such a large dataset (>1200 wines).

In this study, we applied the MWP to a large dataset of 1,288 Gamay wines from the 2022 and 2023 vintages, collected across the ten Beaujolais crus. To address these questions, we applied multivariate statistical methods and machine learning approaches to the mineral datasets. The aim was to evaluate whether the elemental profiles of wines could (i) differentiate vintages, (ii) discriminate among the ten Beaujolais crus despite their geographical proximity, and (iii) explore associations between soil composition and mineral signatures as a mechanistic basis for terroir classification.We hypothesize that the mineral fingerprints of wines provide a robust and reproducible tool for authentication at both the vintage and cru levels, thereby contributing to terroir characterization and offering practical applications for fraud prevention and quality control in Beaujolais wines.

## Materials and methods

### 1. Reagents and materials

All reagents employed were of analytical grade. Ultrapure water with a resistivity of 18.2 megohms and nitric acid of Suprapur® grade (69% (v/v), Roth) were utilized for both sample dilution and standard preparation. Certified metal-free tubes were exclusively utilized for the collection and preparation of both samples and standards.

The semiquantitative calibration standard was meticulously prepared by diluting the multi-element standard, procured from VWR, with a solution containing 100 milligrams per liter of Al, Ag, As, B, Ba, Be, Bi, Ca, Cd, Co, Cr, Cu, Fe, K, Li, Mg, Mn, Mo, Na, Ni, Pb, Sb, Se, Sr, Ti, Tl, V, and Zn in 5% (v/v) HNO3. To optimize the signal intensities of the ICP-MS, an ICP-MS tuning solution was employed, comprising 1 microgram per liter of Ce, Co, Li, Tl, and Y in 2% (v/v) HNO3, sourced from Agilent Technologies.

Validation and control of the semi-quantitative calibration were ensured through the repeated analysis of an in-house reference wine (internal QC material). Additionally, the indium solution utilized as an internal standard was prepared by diluting the 1,000 milligrams per liter indium standard in 4% (v/v) HNO3, obtained from SCP Science.

An in-house reference wine was prepared as a single homogenized batch (blend of Chardonnay, Pinot Noir, and Syrah wines), aliquoted into metal-free tubes, and stored under controlled conditions to minimize evaporation and contamination. This reference wine was used as an **internal QC material** to monitor instrumental stability and batch-to-batch comparability over the full acquisition period. It is not a certified reference material (CRM) and therefore does not provide element-wise trueness; its purpose is to track repeatability and drift.

## 2. Sample preparation

A total of 1,288 Gamay wines were analyzed: 2022 (n=691) and 2023 (n=597), distributed across the ten Beaujolais crus. Approximately 30 mL of wine was collected in metal-free tubes.Samples underwent a primary dilution of 1:3 in 1% $HNO_3$ containing 10 µg/L indium, ensuring stability of elemental composition and preventing precipitation [22]. A secondary dilution of 1:5 in 1% $HNO_3$ was performed immediately prior to analysis, giving a final dilution of 1:15. This minimized matrix effects (alcohol, organics) while allowing accurate trace element determination.
To ensure reproducibility while respecting contractual confidentiality constraints related to our collaboration with M&Wine, we now provide in the Supplementary Information: (i) sample counts per cru and vintage, (ii) descriptive statistics for each of the 41 elements by cru (median, IQR, and range). These additions enable independent assessment of elemental trends and model inputs while preserving sensitive metadata.

## 3. ICP-MS analysis

Analyses were conducted from June 2022 to February 2024 using an Agilent 7850 quadrupole ICP-MS with an SPS 4 autosampler and micromist nebulizer. The collision cell operated in He mode (5 mL/min) to reduce polyatomic interferences. Daily tuning was performed with a calibration solution. Operating parameters were: RF power 1550 W; plasma gas 15 L/min; carrier gas 1 L/min; auxiliary gas 1 L/min.
The Mineral Wine Profile (MWP) comprised 41 elements. Semi-quantitative analysis was performed with 100 sweeps per replicate. Elements not included in the calibration standard were estimated from a reference response table. Analytical throughput exceeded 200 samples/day. Quality control included analysis of a reference wine at the beginning, midpoint, and end of each sequence, blanks and calibration checks every 40 samples, and monitoring of indium in each run. The reference wine was used as an internal QC to monitor repeatability and drift across sequences and does not substitute for a certified reference material (CRM) for trueness.

## 4. Analytical performance (semi-quantitative ICP-MS) and QA/QC

The Mineral Wine Profile (MWP) is designed for high-throughput, comparative multi-element fingerprinting in wine rather than regulatory-grade absolute quantification for each element. Analytical performance was characterized to support fitness-for-purpose of the elemental data used in chemometrics. Linearity was assessed over multiple calibration levels and showed coefficients of determination typically close to unity across representative elements (Table S1). Limits of detection and quantification were calculated from 20 measurements of a 1% $HNO_3$ blank using LOD = μ_blank + 5σ_blank and LOQ = μ_blank + 10σ_blank (Table S2). Repeatability was assessed by 20 successive analyses of the same wine over three days; element-wise means, standard deviations, and RSD% are reported (Table S3).

The present MWP workflow is semi-quantitative and primarily intended for comparative fingerprinting. Element-wise trueness against a wine-matrix certified reference material (CRM) was not established for the full 41-element panel, as fit-for-purpose multi-element wine CRMs are not readily available for our full element panel and concentration ranges. To mitigate matrix effects, samples were diluted to a final 1:15 in 1% $HNO_3$ and an internal standard (In) was used to monitor signal stability. Therefore, the analytical validation presented here focuses on linearity, LOD/LOQ, and repeatability (Tables S1–S3), and conclusions are formulated accordingly.

**Note on matrix effects and method scope.** A two-step dilution (final 1:15 in 1% $HNO_3$) was used to reduce organic load and mitigate matrix-related signal suppression/enhancement. The reported LOD/LOQ and repeatability provide the key uncertainty context for interpretation of both concentration ranges and classification performance.

## 5. Statistical Analysis

All analyses were performed in R (v4.3.2) [23]. Visualization used ggplot2 [24] and plotly.

- Principal Component Analysis (PCA) was performed with the ropls package [25], based on the correlation matrix. The number of PCs was selected according to variance explained (>70%) and scree plot inspection. PCA was used for exploratory data visualization prior to supervised classification.
- Random Forest (RF) classification was performed with the biosigner package [26]. Forests consisted of 1000 trees, with the dataset partitioned into 70% calibration / 30% validation, repeated 1000 times. RF inherently handles imbalanced datasets through bootstrapping, and class weights were verified not to bias the results.
- Mixture Discriminant Analysis (MDA) was performed with the mda package [27]. As MDA is more sensitive to class imbalance, balanced resampling (equalized class sizes by random stratification) was applied before each iteration.

Performance was evaluated using classification accuracy (average of 1000 resampling iterations) and confusion matrices. Misclassifications were analyzed in relation to geographical proximity and soil composition of the crus.

## Results and discussion

The overall mineral content of wine is low, representing less than 0.5% of total mass. Concentrations span eight orders of magnitude, from major cations (K, Ca, Mg, Na: 10–1000 mg/L) to trace elements (Fe, Cu, Zn, Mn: 0.1–10 mg/L) and ultra-traces such as lanthanides (<1 µg/L). In this study, 41 elements were profiled in 1288 Gamay wines: 691 from 2022 and 597 from 2023, covering the ten Beaujolais crus.

Elemental distributions confirmed the predominance of K, Ca, Mg, and Na, while transition metals such as Fe, Mn, Zn, and Cu contributed at mg/L or sub-mg/L levels. Lanthanides (La, Ce, Nd) and some heavy metals (Cd, Pb, Tl) were detected in the ng/L range. These ranges are consistent with previous ICP-MS studies on European wines [9], [20], [21].

Principal Component Analysis (PCA) performed on the 41 elements separated the two vintages along PC1 and PC2, which together explained ~70% of variance (Fig. 3). Despite partial overlap of individual bottles, the centroid positions of 2022 and 2023 samples were distinct.

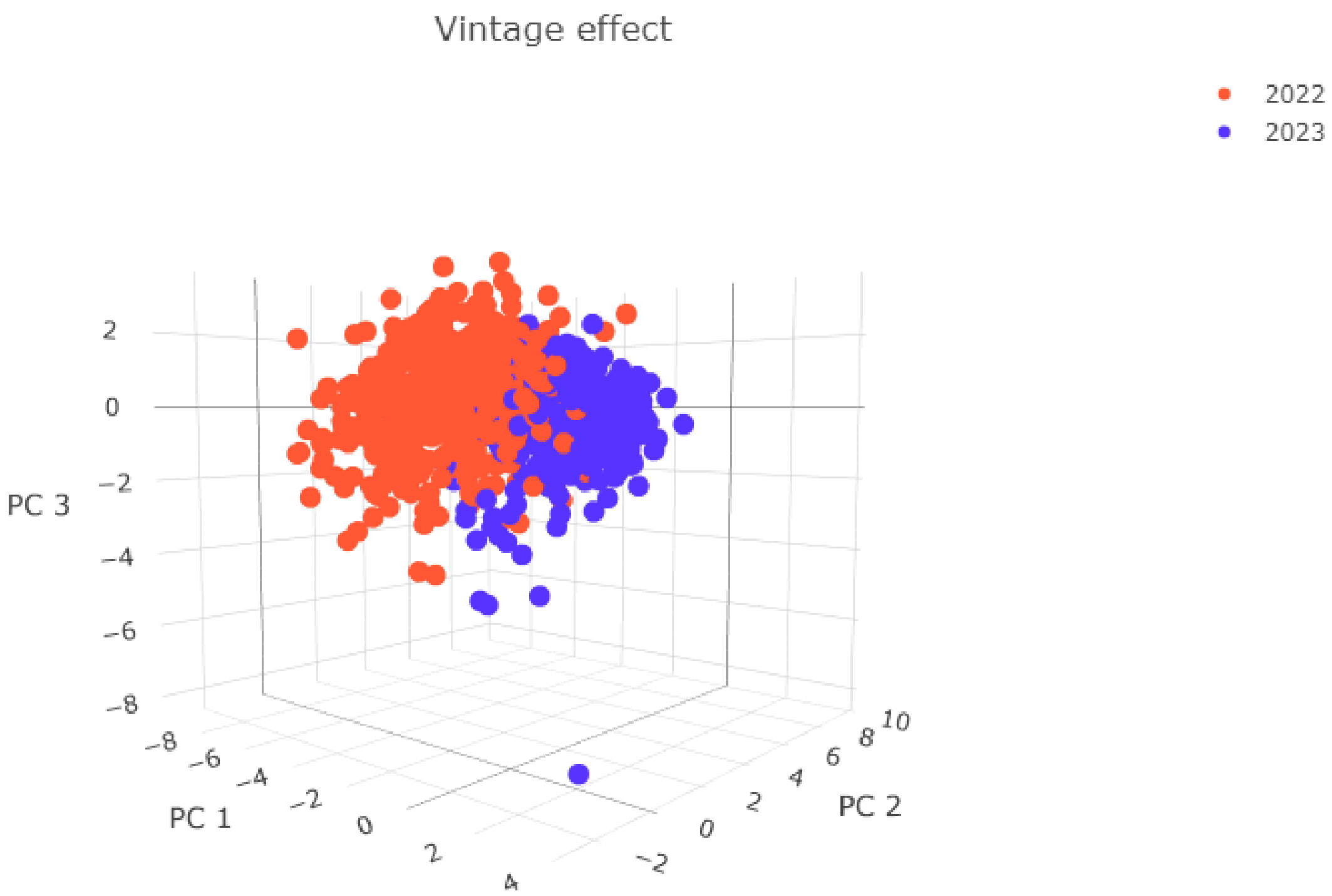

**Fig. 3: Projection of PCA scores from mineral data on to the first three principal components, which represent approximately 70% of the initial variance.**

Random Forest (RF) classification yielded an average accuracy of 98% (Table 1), based on 1000 iterations of 70/30 calibration/validation splits. Misclassification was symmetrical and limited to ~2%, with no evidence of bias toward a particular vintage.

This confirms that vintage effects are stronger than intra-cru differences, consistent with the influence of annual climatic conditions on vine physiology and mineral uptake. Wu et al. (2021)[21] similarly demonstrated that year-to-year variation can significantly modify elemental signatures across French regions.

| Vintage | 2022 | 2023 |
|---|---|---|
| 2022 | 98% | 2% |
| 2023 | 2% | 98% |

**Table 1: Confusion matrix resulting from the cross-validation of Random Forest for vintage discrimination. Randomly selected cross-validation performed on 30% of the data and repeated 1000 times randomly**

When median values of each cru were analyzed by PCA (Fig. 4), separation of the ten crus was observed, explaining ~71% of total variance. Spatial clustering broadly reflected geography:

- Northern crus (Saint-Amour, Juliénas, Chénas, Moulin-à-Vent) grouped together.
- Central crus (Chiroubles, Fleurie, Morgon, Régnié) formed a second cluster.
- Southern crus (Côte de Brouilly, Brouilly) occupied distinct positions.

This gradient suggests that mineral profiles follow soil distributions across the Beaujolais area.

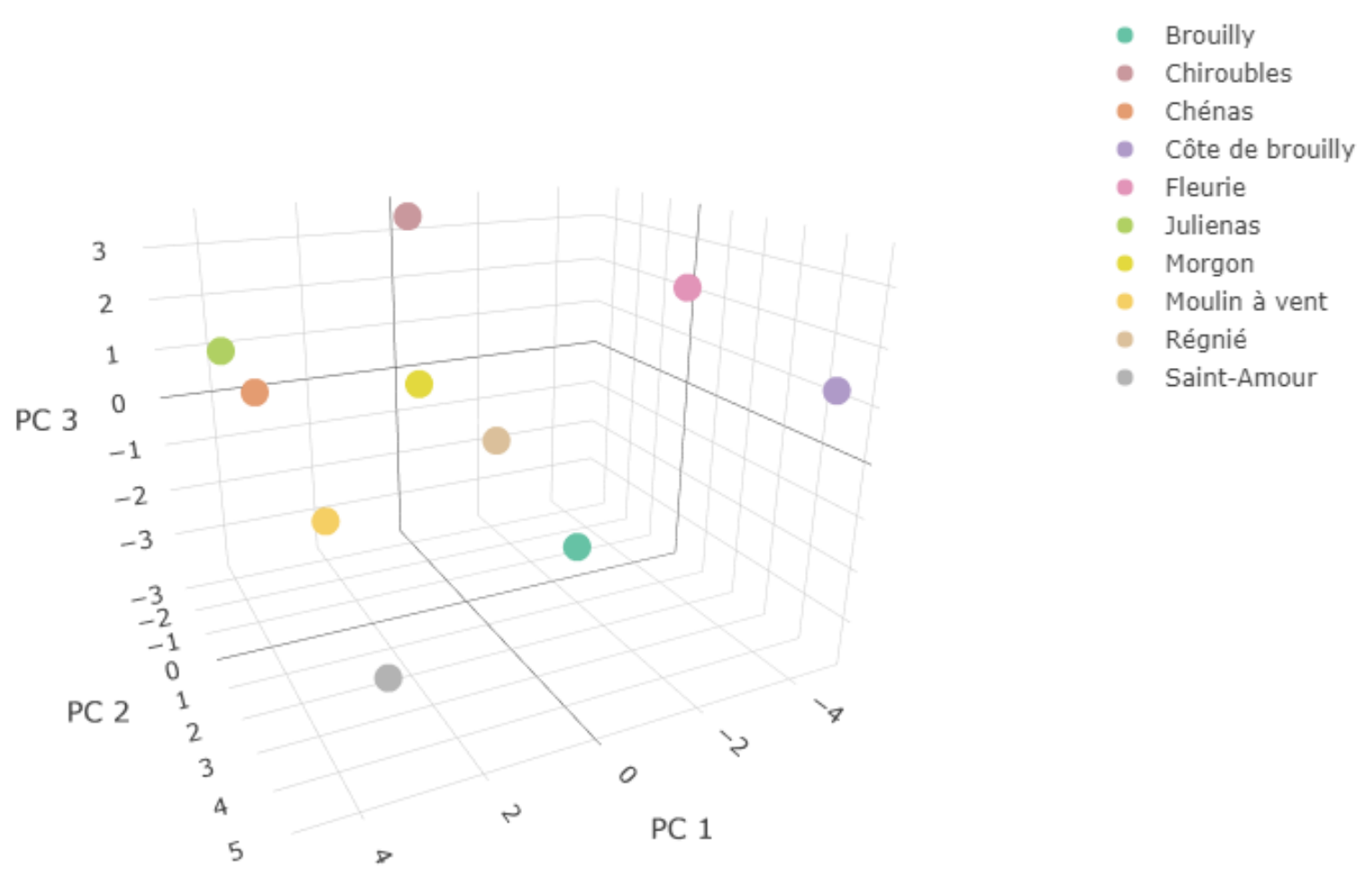

**Fig. 4: Projection of PCA scores from the median of each appellation on to the first three principal components, which represent approximately 71% of the initial variance.**

In PCA, the first two components explained 52% of total variance (PC1 = 34%, PC2 = 18%), while the first three components together accounted for ~71% of the variance. This indicates that most of the variability in elemental profiles can be visualized in two or three dimensions.

Correlation analyses (Fig. 5) revealed specific relationships between soil texture and wine elemental composition:

- Clay-rich soils: enrichment in Fe, Co, Rb, Cr; depletion in Mn, Ca, Sr, Ba.
- Sandy soils: enrichment in Mn, Ca, Sr, Ba; depletion in Fe, Ti, Cr, Rb.
- Silty soils: enrichment in Cs, P, Sn, Cr, Rb; depletion in Mn, Ca, Sr, Ba.

It should be noted that the PCA shown in Fig. 4 is computed from cru-level median elemental values, which intentionally reduces within-cru variability. This choice facilitates visualization of inter-cru trends but may visually enhance separation compared with bottle-level PCA. To document within-cru dispersion, bottle-level descriptive statistics are provided in the Supplementary Information (Tables S4–S5). mechanistic explanationThese findings are consistent with pedological data: clay soils enhance cation exchange capacity and micronutrient retention, while sandy soils favor leaching and reduced metal accumulation. Similar associations have been reported in Bordeaux and Burgundy [4], [11].

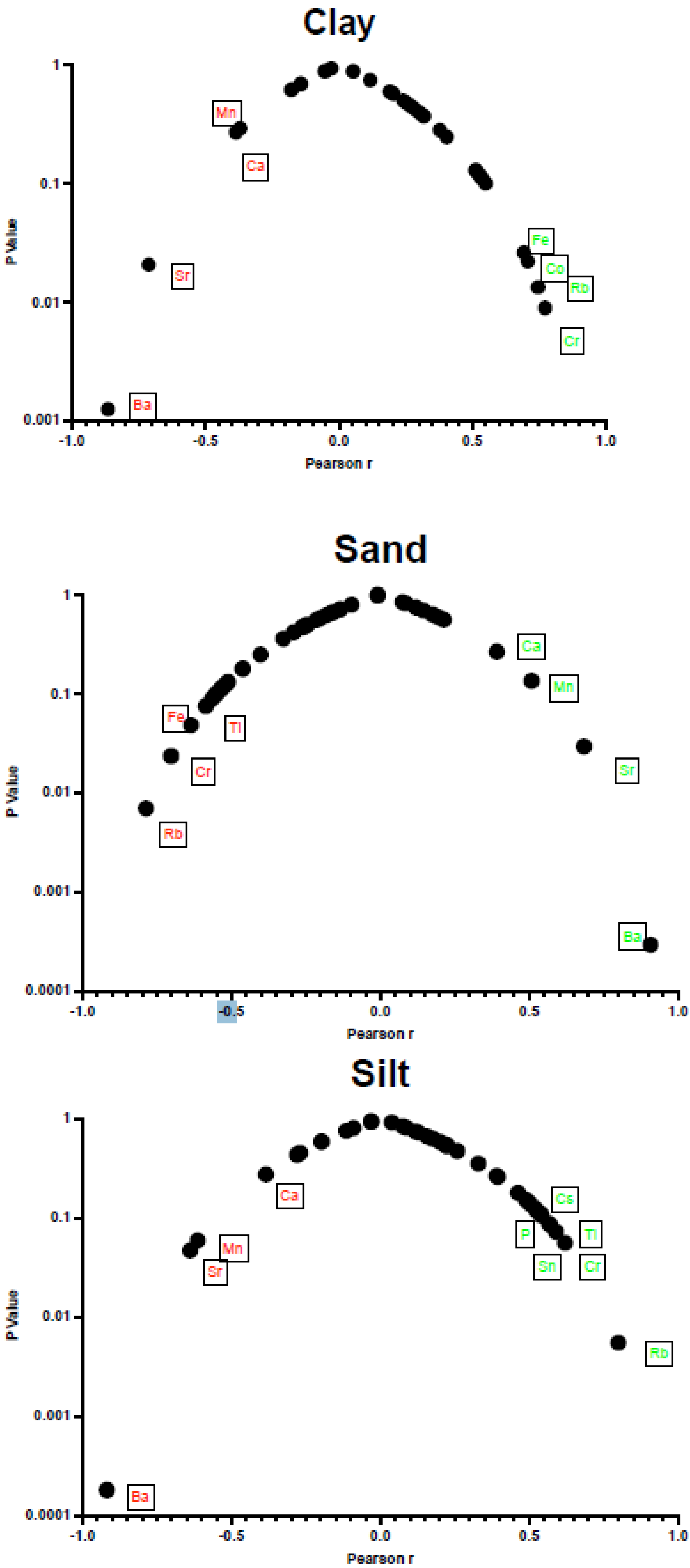

**Fig. 5: The correlation between the medians of metallic and mineral elements contained within a wine bottle and the type of soil (clay, sand, or silt)**

To further investigate the mechanistic link between soil composition and wine mineral profiles, Pearson correlation analyses were performed between the relative abundance of soil fractions (clay, sand, silt) and the concentrations of the 41 measured elements. The results are summarized in Fig. 5, which displays correlation coefficients (r) against their associated p-values for each soil type.

For clay soils, positive correlations were observed with Fe, Rb, Co, and Cr ($r > 0.5$, $p < 0.01$), indicating that wines from clay-rich terroirs tend to be enriched in transition metals and alkali metals associated with higher cation-exchange capacity. Conversely, Mn, Ca, Sr, and Ba showed significant negative correlations with clay content ($r < –0.5$, $p < 0.01$). This suggests that clay-rich soils reduce the mobility of these alkaline earth elements, limiting their transfer into the grape must.

For sandy soils, the trend was reversed. Mn, Ca, Sr, and Ba displayed strong positive correlations ($r > 0.5$, $p < 0.01$), while Fe, Cr, Rb, and Ti were negatively correlated ($r < –0.5$, $p < 0.01$). Sandy soils are more permeable and prone to leaching, leading to higher uptake of mobile cations such as $Ca^{2+}$ and $Sr^{2+}$, whereas less mobile transition metals are less efficiently transferred to the vines.

For silty soils, enrichment patterns differed again. Wines from silty terroirs showed positive correlations with Cs, P, Sn, Cr, and Rb ($r > 0.4$, $p < 0.05$), while Mn, Ca, Sr, and Ba were significantly negatively correlated ($r < –0.4$, $p < 0.05$). These results suggest that silt retains certain trace elements (Cs, Sn) and enhances their availability, while restricting the uptake of more common alkaline earths.

Overall, this analysis demonstrates that soil texture exerts a strong influence on elemental wine composition, in line with known pedological mechanisms [4], [21]. Clay soils tend to concentrate transition metals, sandy soils favor alkaline earth uptake, and silty soils promote specific trace element accumulation. This provides a mechanistic explanation for the terroir differentiation observed in the PCA and MDA models, and supports the interpretation that mineral fingerprints are not random but reflect underlying soil–vine interactions.

While the observed element–soil texture correlations are consistent with known pedological trends, they remain indirect and should be interpreted as associative rather than strictly causal in the absence of paired soil–vine–wine measurements from the same parcels. Elemental fingerprints in wine can reflect terroir-related signals, but they may also incorporate contributions from viticultural practices (e.g., fertilization regimes, foliar treatments), winemaking inputs (e.g., bentonite fining, yeast nutrients), and contact materials (equipment, storage and bottling components). Although the observed associations between soil texture fractions and elemental medians are consistent with pedological trends, they remain indirect and should be interpreted as associative rather than strictly causal in the absence of paired parcel-level soil–vine–wine measurements. Accordingly, we frame our conclusions primarily in terms of authentication-oriented fingerprinting rather than mechanistic proof of soil transfer.The differentiation among the ten Beaujolais crus was subsequently assessed using Mixture Discriminant Analysis (MDA). The classification model achieved an average accuracy of 92%, indicating that the majority of wines could be correctly assigned to their appellation based solely on their mineral fingerprints (Table 2).

However, beyond overall accuracy, a detailed examination of misclassifications provides additional insights. Fleurie, for example, showed one of the highest classification rates (85%), yet approximately 10% of its wines were misclassified as Morgon. Similarly, Chiroubles wines were correctly classified in 87% of cases, but 13% were predicted as Morgon. These errors can be explained by the geographical proximity and shared soil structures of these crus: Fleurie, Chiroubles, and Morgon form a contiguous zone in central Beaujolais, with overlapping granite-derived soils and similar altitudes. Such physical and pedological similarities make it statistically more difficult to distinguish these appellations using elemental data alone.

This pattern was consistent across the dataset: most misclassifications occurred between neighboring crus (e.g., Côte de Brouilly vs Brouilly, Moulin-à-Vent vs Chénas). Importantly, these errors rarely involved distant crus with contrasting geology. This indicates that the classification model is not random but strongly reflects the underlying terroir structure.

When misclassifications were aggregated at a broader spatial scale, 95% of wines were assigned to the correct geographical sub-region of Beaujolais. This suggests that elemental profiles capture meaningful geographical gradients, even when boundaries between adjacent crus are blurred.

| predicted | Brouilly | Chénas | Chiroubles | Côte de Brouilly | Fleurie | Juliénas | Morgon | Moulin à Vent | Régnié | Saint-Amour |
|---|---|---|---|---|---|---|---|---|---|---|
| Brouilly | 86% | 0% | 0% | 0% | 0% | 3% | 3% | 7% | 2% | 0% |
| Chénas | 0% | 100% | 0% | 0% | 0% | 0% | 0% | 0% | 0% | 0% |
| Chiroubles | 0% | 0% | 87% | 0% | 0% | 0% | 0% | 0% | 0% | 0% |
| Côte de Brouilly | 6% | 0% | 0% | 91% | 5% | 0% | 3% | 0% | 0% | 0% |
| Fleurie | 0% | 0% | 0% | 4% | 85% | 0% | 1% | 0% | 0% | 0% |
| Juliénas | 1% | 0% | 0% | 0% | 0% | 94% | 5% | 0% | 0% | 2% |
| Morgon | 4% | 0% | 13% | 0% | 10% | 3% | 88% | 5% | 4% | 0% |
| Moulin à Vent | 0% | 0% | 0% | 4% | 0% | 0% | 0% | 88% | 0% | 0% |
| Régnié | 0% | 0% | 0% | 1% | 0% | 0% | 0% | 0% | 94% | 0% |
| Saint-Amour | 3% | 0% | 0% | 0% | 0% | 0% | 0% | 0% | 0% | 98% |

**Table 2: Confusion matrix resulting from the cross-validation of MDA for the discrimination of the 10 Beaujolais crus. Randomly selected cross-validation performed on 30% of the data and repeated 1000 times randomly. Values represent average classification rates across 1000 resampling iterations.**

To further validate robustness, a one-vs-all classification approach was applied, where each cru was modeled against all other appellations (Table 3). This method reduces the confounding effect of overlaps between neighboring crus and tests whether each appellation carries a distinctive signature relative to the rest. Results showed that over 94% of bottles were correctly classified across all crus, with some appellations reaching even higher accuracy (e.g., Saint-Amour: 98%; Fleurie: 97%). This confirms that despite local overlaps, each cru exhibits a sufficiently unique mineral profile to be distinguished from the others. Across the 1000 iterations, the standard deviation of prediction accuracy was <2%, confirming the high classification performance under repeated resampling and reproducibility of the models.

| predicted | Beaujolais crus | other |
|---|---|---|
| Brouilly | 90% | 10% |
| Chénas | 99% | 1% |
| Chiroubles | 96% | 4% |
| Côte de Brouilly | 89% | 11% |
| Fleurie | 97% | 3% |
| Juliénas | 95% | 5% |
| Morgon | 88% | 12% |
| Moulin à Vent | 91% | 9% |
| Régnié | 95% | 5% |
| Saint Amour | 98% | 2% |

**Table 3: Confusion matrix resulting from the cross-validation of MDA for the discrimination of the 10 Beaujolais crus vs other. Randomly selected cross-validation performed on 30% of the data and repeated 1000 times randomly. Values represent average classification rates across 1000 resampling iterations.**

Taken together, these results demonstrate that the MWP approach provides stable performance across resampling intra-regional authentication:

- At the individual cru level, >9 out of 10 bottles are correctly classified.
- At the broader regional scale, >95% are assigned correctly to their geographical zone.
- Misclassifications are not random, but correspond to logical overlaps between neighboring crus with similar soil and geological features.

These findings are consistent with previous observations in other French regions, where misclassifications also tended to occur between adjacent appellations sharing soil characteristics [11], [21]. Our study extends this evidence by showing that even within a single grape variety and a relatively small geographical area, mineral fingerprints provide sufficient resolution to authenticate terroirs at a fine scale. These levels of accuracy are comparable to, or higher than, those reported in previous ICP-MS studies at larger geographical scales [9], [21], highlighting the capacity of mineral fingerprints to resolve intra-regional differences within a single grape variety.

Beyond its analytical performance, this work also highlights the economic and practical implications of terroir authentication. The Beaujolais crus represent a significantly higher market value compared to generic Beaujolais wines, sometimes reaching prices several times higher. Reliable authentication of origin therefore contributes directly to protecting this added value, strengthening producers' competitiveness, and ensuring consumer confidence. For winegrowers and local cooperatives, the availability of high-throughput mineral profiling offers a practical tool to demonstrate the uniqueness of their products and to reinforce communication strategies based on terroir. For regulatory agencies, such as AOC control boards, the approach provides an objective and reproducible method to complement traditional certification schemes.

Similar challenges exist in other emblematic wine regions, such as the Douro Valley (Portugal)[6], Rioja (Spain)[28], or Napa Valley (USA)[29], where terroir-driven products also depend on strict authenticity guarantees. Extending mineral profiling combined with machine learning to these contexts could support international harmonization of control methods and provide a scientific basis for protecting geographical indications worldwide. In this sense, the Beaujolais case study not only demonstrates the local relevance of the Mineral Wine Profile (MWP), but also positions it as a scalable model for global viticulture, linking analytical innovation with socio-economic valorization of terroirs.

## Conclusion

This work demonstrates that mineral profiling of wines, using the Mineral Wine Profile (MWP) combined with machine learning, enables reliable authentication of Beaujolais wines at both vintage and cru level.

Key findings include:

- Vintage effects: PCA and RF showed near-perfect discrimination (98% accuracy) between 2022 and 2023.
- Cru discrimination: MDA achieved 92% accuracy overall, with misclassifications largely confined to neighboring crus. One-vs-all classification improved accuracy to >94% for each cru.
- Soil-element association: Element–soil texture associations (e.g., Fe–clay, Mn–sand, Cs–silt) were consistent with terroir-related gradients but remain indirect without paired parcel-level soil analyses; interpretations are therefore phrased as associative rather than strictly mechanistic.Despite occasional misclassifications, >95% of wines were assigned to the correct geographical sub-region. This confirms that even within a compact region, distinct elemental signatures persist and can be exploited for authentication.

Compared to previous studies limited to inter-regional discrimination, our results highlight the potential of MWP as a scalable, reproducible, and high-throughput protocol for fine-scale wine authentication.

Beyond their scientific interest, these results have practical implications for the protection of Appellations d'Origine Contrôlée (AOC). The ability to authenticate wines at both vintage and cru level provides a potential tool for fraud prevention and regulatory control. Moreover, the high-throughput MWP approach is scalable and could be applied to other wine regions or agricultural products requiring origin authentication. Limitations of this study include the absence of paired parcel-level soil measurements and the fact that the workflow is semi-quantitative and primarily intended for comparative fingerprinting; however, sample distribution and descriptive statistics of elemental concentrations are now provided in the Supplementary Information to support reproducibility.Such integration would not only improve the understanding of terroir effects but also enhance the application of mineral fingerprinting in fraud prevention, terroir valorization, and regulatory control.

## Credit authorship contribution statement

**Yohann Clément** : Writing – original draft, Visualization, Methodology, Investigation, Formal analysis, Conceptualization**, Pierre Lantéri:** Writing – original draft, Visualization, Methodology, Investigation, Formal analysis, Conceptualization. **Agnès Agège:** Resources. **Bertrand Chatelet**: Resources, Investigation. **Théodore Tillement:** Resources, Investigation. **Coraline Duroux**: Resources, Investigation, Formal analysis. **Leticia Sarlo:** Writing – review & editing, Resources. **François Lux:** Writing – review & editing, Writing – original draft, Supervision, Project administration, Funding acquisition, Conceptualization. **Olivier Tillement:** Writing – review & editing, Writing – original draft Supervision, Software, Formal analysis, Conceptualization.

## Declaration of competing interest

Several authors are affiliated with M&Wine, which contributed to sample logistics and/or access to the wine dataset used in this study. No funding was received from M&Wine specifically to conduct this work, and the company had no role in study design, data analysis, interpretation, or the decision to publish. The authors declare no other competing financial interests or personal relationships that could have appeared to influence the work reported in this paper.

## Data availability

The sample counts per cru and vintage, and descriptive statistics by cru (median, IQR, range) are provided in the Supplementary Information. Producer- and parcel-level identifiers are not shared due to contractual confidentiality constraints.

## Acknowledgments

We would like to express our gratitude to Victor Gomez and Henri-Laurent Arnould, wine competition organisers, and to Gilles Masson, president of Centre du Rosé, for providing the majority of the samples. We also thank M&Wine for support in sample logistics and dataset access within the framework of an academic–industry collaboration, with no influence on data processing, interpretation, or publication decisions.

## References (APA Style is mandatory)